# A Systematic Methodology for Developing Discrete Event Simulation Models of Software Development Processes

Ioana Rus, Holger Neu, and Jürgen Münch

*Abstract*— So far there have been several efforts for developing software process simulators. However, the approaches for developing the simulators seem to have been ad-hoc and no systematic methodology exists. Since modeling and simulation in support of software development should become more popular (and there are signs that it does), there is a need for migrating modeling from craft to engineering. This article proposes such a systematic method, focused on the development of discrete simulation-based decision models, but extensible to other modeling approaches as well.

## I. INTRODUCTION

Most publications that describe simulation models for software development processes do not address the process followed for developing these models. That leads us to believe that they are developed ad-hoc, in isolation, many from scratch, and that there are no systematic and documented ways of developing them. For system dynamics models, some authors describe their approach for developing the models [4][6]. For discrete event models, no process has been published that describes the development of such a model.

Reflecting on how a software process simulator has been developed and sharing this process with other model developers is very valuable, as it allows approaching simulation in a systematic manner and provides guidance and support to the modeling community. The application of such a methodology will enhance the quality of the resulting model (including its reusability and maintainability) and will reduce development effort and duration.

In this article we present a method for systematically developing discrete event simulation models for software development processes, which is derived from the authors' collective experience from building a few simulators.

We focus on describing the development process of a discrete event simulator, but this can easily be applied to different modeling approaches. For identifying and describing the modeling process, we suggest applying practices and principles from software engineering.

The intended audience for this article is the modeling and simulation community of practice.

This article is organized as follows: Section II presents the activities of the proposed method. For each of the activities, we will present a description and will illustrate it with examples from our own work. Section III contains a summary and conclusions, and Section IV highlights directions for future work.

## II. PROPOSED METHOD

This method considers the development of a new simulation model without reusing or incorporating existing components. If reuse is considered (either incorporating existing components or developing for reuse), then the method has to be changed to address possible reuse of elements.

We believe that the life cycle of a simulation model for long-term use is similar to the one of software, consisting of three main phases: development, deployment, and operation, including maintenance and evolution (after all, a simulation model is a software product, too). The activities within these phases can have different time order and dependencies, therefore, the resulting life cycle can take on different forms, such as waterfall, iterative, or even agile.

The activities throughout the life cycle can be divided in two categories: *engineering* and *management* activities. Since most of our experience is related to the development phase of a simulator, this is what we will focus on in this article.

The engineering activities for model development are:
- Requirements identification and specification for the model to be built;
- Analysis and specification of the modeled process;
- Design of the model;
- Implementation of the model;
- Verification and validation throughout development.

The management activities are:

Ioana Rus is with the Fraunhofer Center for Experimental Software Engineering, College Park, MD, 20740, USA (corresponding author's phone: 301-403-8971; fax: 301-403-8976; e-mail: irus@fc-md.umd.edu).

Holger Neu is with the Fraunhofer Institute Experimental Software Engineering (IESE), Sauerwiesen 6, 67661 Kaiserslautern, Germany (e-mail: Holger.Neu@iese.fraunhofer.de).

Jürgen Münch is with the Fraunhofer Institute for Experimental Software Engineering (IESE) in Kaiserslautern, Germany (e-mail: muench@iese.fraunhofer.de).

- Model development planning and tracking;
- Measurement of the model and of the model development process;
- Risk management – refers to identification and tracking of risk factors, and mitigation of their effects. Some of the risk factors are: changes in customer's requirements, changes in the description of the modeled process, non-availability of data for the quantitative part of the model.

During the life cycle of a simulation model, different roles are involved. In the development phase, mainly the model developer and the customer are involved. We did "pair modeling" for creating the first version of the static process model and influence diagram. This seems very useful because the discussion about the model and the influences is very inspiring.

*A. Simulator requirements identification and specification*

During the requirements activity, the purpose and the usage of the model have to be defined. According to this, the questions that the model will have to answer are determined and so is the data that will be needed in order to answer these questions. Sub-activities of the requirements specification are:

*1) Definition of the goals, questions, and the necessary metrics*

A goal-question-metrics-(GQM) [1][3] based approach for defining the goal and the needed measures seems appropriate. GQM can also be used to define and start an initial measurement program if needed.

The purpose, scope and level of detail for the model is described by the *goal*. The *questions* that the model should help to answer are formulated next. Afterwards, parameters (*metrics*) of the model (outputs) have to be defined that (once their value is known) will answer the questions. Then the model input parameters have to be defined, which are necessary for determining the output values. The input parameters should not be considered as final after the requirements phase; during the analysis phase they will usually change.

*2) Definition of usage scenarios*

Define scenarios ("use cases") for using the model. For example, for answering the question: "How does the effectiveness of inspections affect the cost and schedule of the project?", a corresponding scenario would be: "All input parameters are kept constant and the parameter *inspection effectiveness* is given x different values between (min, max). The simulator is executed until a certain value for the number of defects per KLOC is achieved, and the values for *cost* and *duration* are examined for each of the cases."

For traceability purpose, scenarios should be tracked to the questions they answer (for example, by using a matrix).

*3) Development of test cases*

Test cases can be developed in the requirements phase. They help to verify and validate the model and resulting simulation.

*4) Validation of requirements*

The customer has to be involved in this activity and must agree with the content of the resulting model specification document. Changes can be made, but they have to be documented.

Throughout this article, we will illustrate the outputs of the activities by using some excerpts from a discrete event simulator developed by us at IESE. This model and simulator supports managerial decision-making for planning the system testing phase of software development. The simulator offers the possibility of executing *what-if* scenarios with different values for the parameters that characterize the system testing process and the specific project. By analyzing the outputs of the simulator, its user can visualize predictions of the effects of his/her planning decisions.

*Example for step 1) and step 2):*

- Goal: Develop a model for decision support for planning of the system testing phase, such that trade-offs between duration, cost, and quality (remaining defects) can be analyzed and the most suitable process planning alternative can be selected, in the context of organization x.

- Questions to be answered:
  Q1: When to stop testing in order to reach a specified quality (number of defects expected to remain in the delivered software)?
  Q2: If the delivery date is fixed, what will be the quality of the product, if delivered at that time, and what will be the cost of the project?
  Q3: If the cost is fixed, when will the product have to be delivered and what will be its quality at that point?
  Q4: Should regression testing be performed? To what extent?

- Output parameters:
  O1: Cost of testing (computed from the effort) [$] for cost or [staff hours] for effort
  O2: Duration of testing [hours] or [days]
  O3: Quality of delivered software [number of defects per K line of code]

- (Some of the) Input parameters:
  I1: Numbers of requirements to be tested
  I2: Indicator of the "size" of each software requirement (in terms of software modules (components) that implement that requirement and their "complexity/difficulty" factor)
  I3: For each software module, its "complexity/difficulty" factor
  I4: Number of resources (test-beds and people) needed
  I5: Number of resources (test-beds and people) available
  I6: Effectiveness of test cases (historic parameter that

gives an indication about how many defects are expected to be discovered by a test case)

- Scenarios:
  S1: For a chosen value of the desired quality parameter, and for the fixed values of the other inputs, the simulator is run once until it stops (the desired quality is achieved) and the values of cost and duration outputs are examined.
  S2: The simulator is run for a simulation duration corresponding to the chosen value of the target duration parameter, and for the fixed values of the other inputs. The values of cost and quality outputs are examined.
  S3: For a chosen value of the desired cost parameter, and for the fixed values of the other inputs, the simulator is run once until it stops (the cost limit is reached) and the values of quality and duration are examined.
  S4: For a chosen value of the desired quality parameter, and for the fixed values of the other inputs, the simulator is run once until it stops (the desired quality is achieved) and the values of cost and duration outputs are examined according to the variation of the extent of regression testing.
  S5: The simulator is run for a simulation duration corresponding to the chosen value of the target duration parameter, and for the fixed values of the other inputs, and the values of cost and quality outputs are examined according to the variation of the extent of regression testing.
  S6: For a chosen value of the desired cost parameter, and for the fixed values of the other inputs, the simulator is run once until it stops (the cost limit is reached) and the values of quality and duration are examined according to the variation of the extent of regression testing.

### B. Process analysis and specification

The understanding, specification and analysis of the process that is to be modeled is one of the most important activities during the development of a simulation model.

We divide process analysis and specification into four sub-activities, as shown in Figure 1: Analysis and creation of a static process model (a), creation of the influence diagram for describing the relations between parameters of the process (b), collection and analysis of empirical data for deriving the quantitative relationships (c), and quantification of the relations (d). Figure 1 sketches the product flow of this activity, i.e., it describes which artifacts (document symbol) are used or created by each task (arrowed circle symbol).

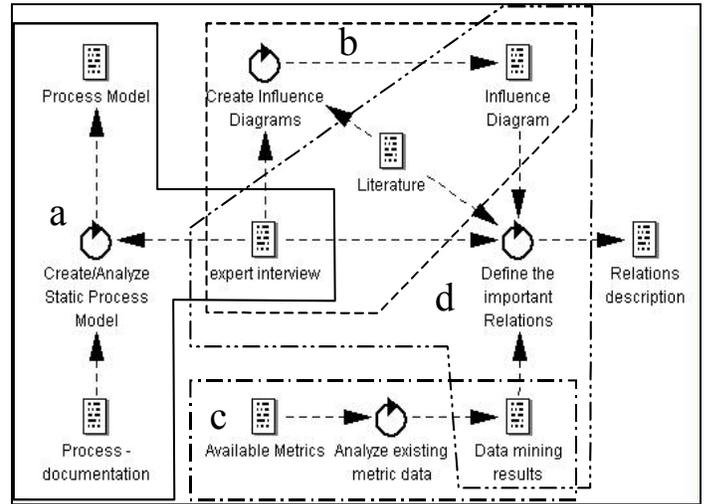

**Figure 1.** Process analysis and specification.

*1) Analysis and creation of a static process model*

The software process to be modeled needs to be first understood and documented. This requires that the representations (abstractions) of the process should be intuitive enough to be understood by the customer and to constitute a communication vehicle between modeler and customer. These representations lead to common definition and understanding of the object of modeling (i.e., the software process) and to a refinement of the problem to be modeled (initially formulated in the requirements specification activity). As input and sources of information for this activity, process documents are possibly supplemented with interviews with people involved in the process (or with the process "owner"). The created process model describes the artifacts used, the processes or activities performed and the roles and tools involved. The process model shows which activities transform which artifacts and how information flows through the model. In our work, to support this activity, we used *Spearmint* as a modeling tool [2].

*2) Creation of the influence diagram for describing the relations between parameters of the process*

For documenting the relationships between process parameters, we use influence diagrams. Influence factors are typically factors that change the result or behavior of other project parameters. The relations between influencing and influenced parameters are represented in an influence diagram by arrows and + or -, depending on whether variation of the factors occurs in the same way or on opposite ways.

When we draw the influence diagram, we should have in mind what the inputs and outputs we identified in the requirements phase are. These inputs and, especially, the outputs have to be captured in the influence diagrams.

Figure 2 presents a small excerpt of a static process model and a corresponding influence diagram.

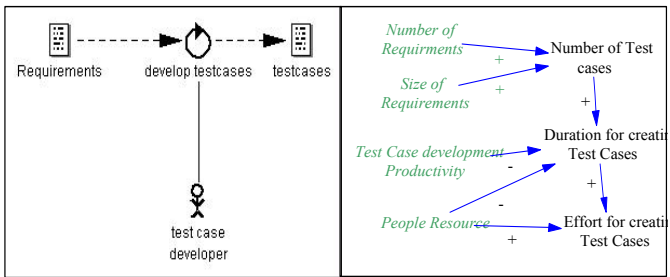

**Figure 2.** Static process model (left) and influence diagram (right).

The influence diagrams that we developed in this step were later refined during design and especially during implementation, driven by a better understanding of what is really needed for implementation.

*3) Collection and analysis of empirical data for deriving the quantitative relationships,*

For identifying the quantitative relations between process parameters, we need to identify which data/metrics need to be collected and analyzed. Usually, not all required data is available, and additional metrics from the target organization should be collected. In this case, developing a process model can help to shape and focus the measurement program of a project.

*4) Quantification of the relations*

This is the task that is probably the hardest part of the analysis. Here we have to quantify the relations and distinguish parameter types as follows:
- Calibration parameters: for calibrating the model according to the organization, like productivity values, learning, skills, and number of developers;
- Project-specific input: for representing a specific project like number of test cases, modules, and size of the tasks;
- Variable parameters: these are the parameters that are changed to analyze the results of the output variables. In general these can be the same as the calibration parameters. The scenario from the requirements or new scenarios determine the variable parameters during the model life cycle.

The distinction between these parameters is often not easy and shifts, especially, for calibration and variable parameters depending on the scenarios that are addressed by the model.

The mathematical quantification is done in this step. Depending on the availability of historical metric data, sophisticated data mining methods might be used to determine these relations. Otherwise, interviews with customers, experts, or literature sources have to be used.

The outputs of the process analysis and specification phase are models (static model, influence diagrams, and relations) of the software development process and parameters that have to be simulated, measures (metrics) that need to be received from the real process, and a description of all the assumptions and decisions that are made during the analysis. The latter is useful for documenting the model for later maintenance and evolution.

The artifacts created during process analysis and specification have two distinct properties: the level of static or dynamic behavior they capture, and the quantitative nature of the information they represent. Figure 3 shows the values for these properties for the static process model (which is static and qualitative), the influence diagram (static and qualitative), and the end product simulator, which is quantitative and dynamic.

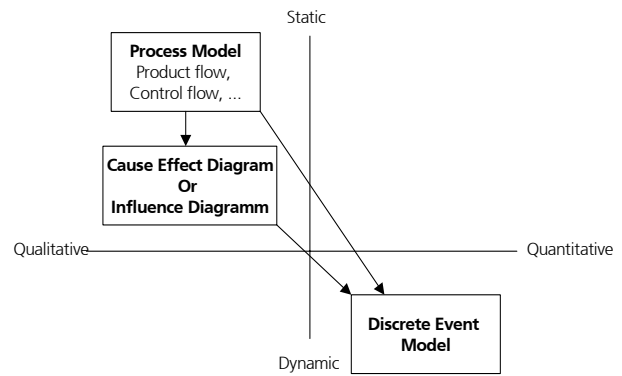

**Figure 3.** Properties of different modeling artifacts.

Throughout this activity, several verification and validation steps must be performed:
- The static model has to be
  a) validated against the requirements (to make sure it is within the scope, and also that it is complete for the goal stated in the requirements specs.);
  b) validated against the real process (here the customer should be involved).
- The parameters in the influence diagram must be checked against
  a) the metrics (that they are consistent)
  b) the input and output variables of the requirement specification.
- The relations must be checked against the influence diagrams for completeness and consistency. All the factors of the influence diagram that influence the result have to be represented, for instance, in the equation of the relation.
- The influence diagrams (possibly also the relations) must be validated by the customer with respect to their conformance to real influences.
- The units of measurement for all parameters should be checked.

*C. Design*

During this activity, the modeler develops the design of the model, which is independent of the implementation environment. The design is divided into different levels of detail, the high-level design and the detailed design.

*1) High Level Design*

In the high level design the surrounding infrastructure is defined. Also, the basic mechanisms describing how the input and output data is managed and represented is defined, if necessary. Usually, a model's design comprises components like a database or a spreadsheet, a visualization component, and the simulation model itself together with the information flows between them. Figure 4 shows the high level design for our system testing simulation (STS) model. The model has two main modules, the *Development module,* which models the development of the software, and the *STS Testing module*, which models the system testing of the software. These two modules interact through the flow of items such as *Code*, *Resources*, and *Defects*. The whole model has GUI interfaces, both for input and for output. Through the input interface, the user of the simulator will provide the values of the input parameters, which will be saved in a database and then fed into the simulator. The outputs of the simulator are saved in the database and from there are then used by a visualization software and displayed in a friendly format to the user of the model.

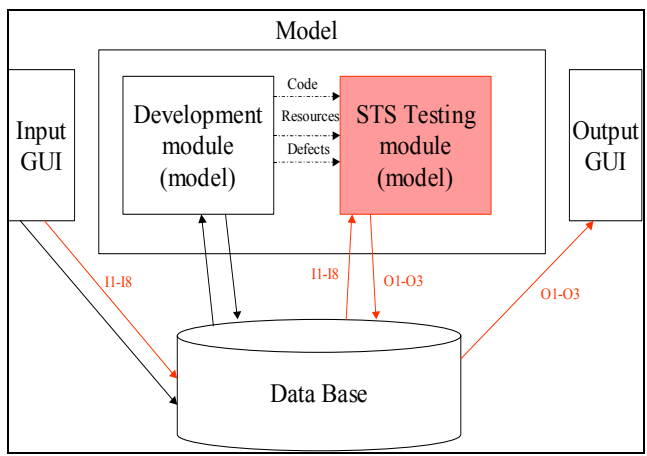

**Figure 4.** High level design (excerpt).

The high level design is the latest point in time for deciding which type of simulation approach to use (e.g., system dynamics, discrete event, or anything else).

*2) Detailed Design*

In the detailed design, the "low level design" of the simulation model is created. The designer has to make several decisions, such as:

- Which activities do we have to model? (what is the level of granularity for the model)
- What items should be represented?
- What are the attributes of these items that should be captured?

Additionally, the designer has to define the flow of the different items (if more than one type of items is defined, also the combination of items). For doing the detailed design, we looked at the static process model (modeled with Spearmint) for identifying the activities, and items, and at the influence diagrams for identifying the items' attributes.

The outcome of this activity is a detailed design of the model. In Figure 5 we show such a design, using a UML-like notation. The *Activity* object in Figure 5 models an activity in a software process. Its attributes are: duration, cost, effort, inputs, outputs, resources type and number (both people and equipment resources). The *Resource* object corresponds to the real world resources needed to perform an activity. An example of instances of the *People_Resource* sub-class of *Resource* would be *Developer* or *Tester*. The human resources are characterized by their productivity, represented as the attribute *Productivity*. The object *Artifacts* captures the code (*SW_Artifacts*) and test cases (*Testing_Artifact*). The code has as attributes its size, complexity, and number of defects, while the test cases have an attribute related to their capability of detecting the defects in code (*Effectiveness*).

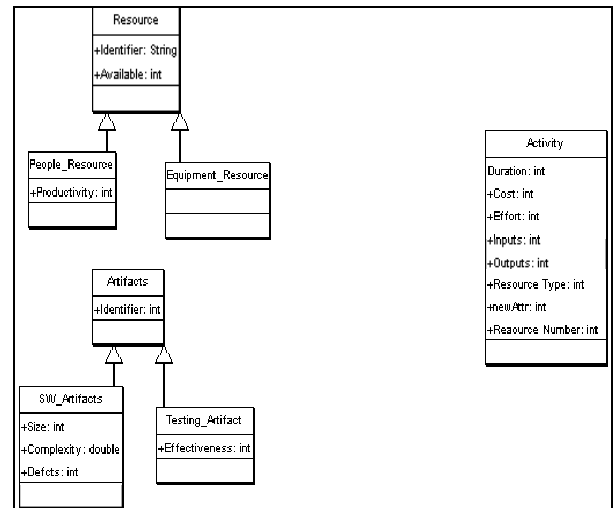

**Figure 5.** Example of detailed design objects.

In the design phase, verification and validation activities must be performed, for example to check consistency between high and low level design and also against process description (static, influence diagrams, and relations) and model requirements.

**D. Implementation**

During implementation, all the information and the design decisions are transferred into the simulation model. The documents from the process specification and analysis and the design are necessary as inputs.

This activity in the development process depends heavily on the used simulation tool or language used and is very similar to the implementation activity for a conventional software product.

Figure 6 shows an excerpt of our discrete event model developed in *Extend* V5.

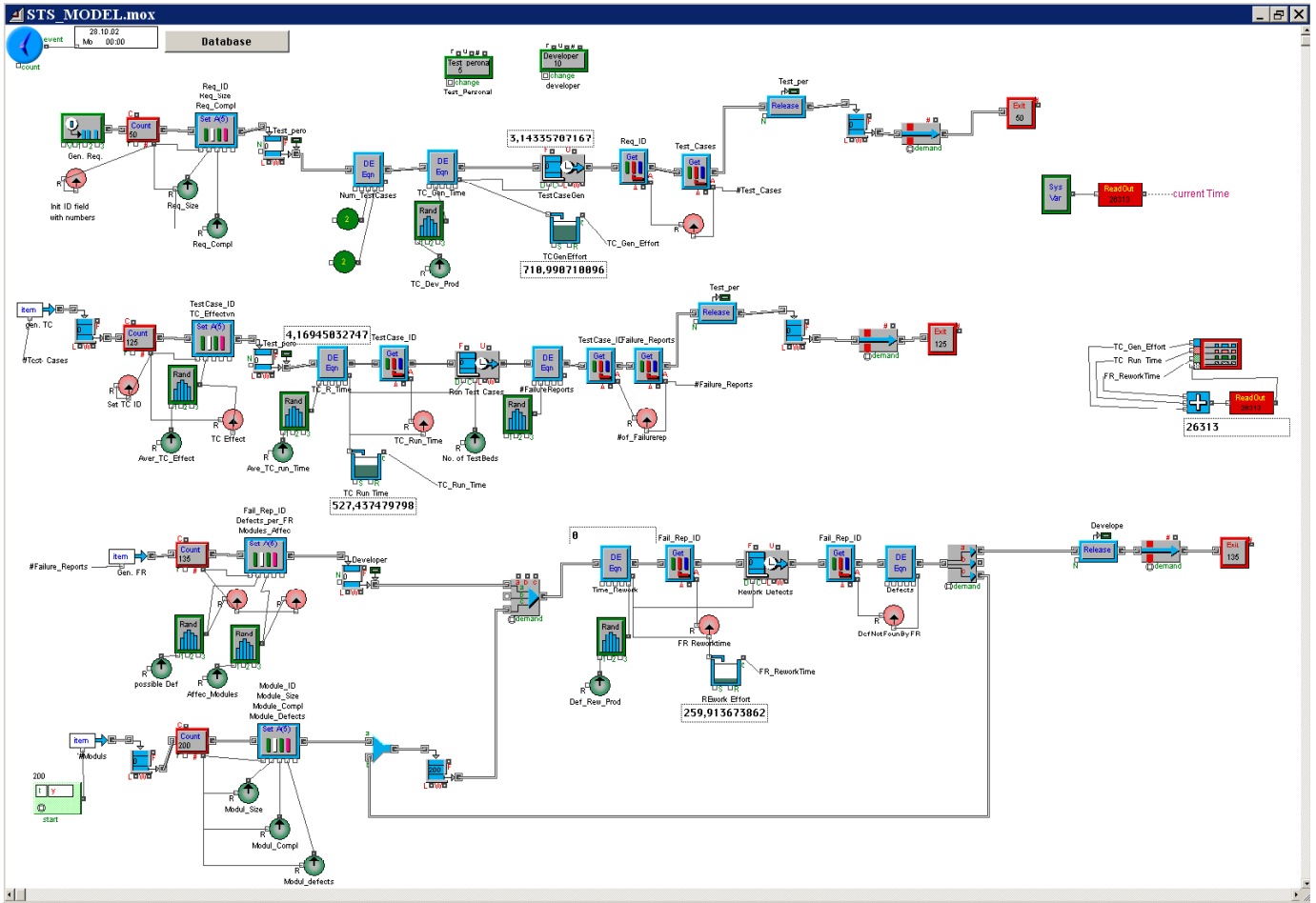

Figure 6: Simulation model.

*A. Validation and verification (model testing)*

Besides the validation and verification that occur throughout the development, the implemented model and simulator must be checked to see whether it is suitable for the purpose and the problems it should address. Also, the model will be checked against the reality (here the customer has to be involved).

Throughout the simulator development process, verification and validation (usually implemented by reviews) must be performed after each activity. Also, traceability between different products created during the simulator's development must be maintained, thus enabling the modeler to easily go back to correct or add elements during model development or evolution. Implementation decisions must be documented for future understanding of the current implementation and for facilitating model evolution.

### III. SUMMARY AND CONCLUSIONS

We presented a systematic process for developing discrete event executable models. However, this process is not restricted to this modeling approach and can be applied as well to developing other types of models, such as continuous (system dynamics) models.

### IV. FUTURE WORK

Future work comprises the systematic integration of further activities (such as model verification and validation) into the method. Additionally, the other phases (deployment, maintenance and evolution) that are briefly sketched below, have to be refined:

In the *deployment phase,* the model is calibrated and given to the customer. This includes several activities that have to be performed. Also, the target environment has to be analyzed to see whether something special is needed. In the calibration step, the model is calibrated with the data of the target organization. The data has to be evaluated using measurement data, expert interviews, data taken from the literature, or applying data mining techniques. During the installation step, the model is installed at the customer's site. The installation depends on the software used to create the simulation model. If a runtime environment for the simulation model is available and all the data is contained in the model, an installation of this runtime environment can be sufficient. If the model was created using special tools and the model needs these tools for being executed, the tools also have to be installed. In order to use the model efficiently, the user needs training, or at least advice on how to use the model. He also has to know the limitations of the model.

In the *maintenance and evolution phase,* a model can be changed by: 1) changing values for the calibration parameters; 2) changing relations (functions, equations); 3) changing the structure of the model. These levels require different effort and modeling skills for change (increasing from 1 to 3).

Possible extensions of the method could be the addition of reuse concepts and the combination of simulation and real experiments. Research questions in this area might be: How to reuse artifacts and knowledge during the simulation modeling process (e.g., reuse of influence diagrams, reuse of parts of the simulation model)? What are appropriate interfaces of a reusable model/influence diagram? What could a framework for reusing simulation models look like? How should real experiments be designed in order to support simulation modeling and to benefit from simulation? How to combine experimentation with simulation for advancing empirical studies in software engineering?

We expect this proposed method to be improved by contributions from the experience of other model developers and users.

ACKNOWLEDGMENT

This work was supported in part by the German Federal Ministry of Education and Research (SEV Project) and the Stiftung Rheinland-Pfalz für Innovation (ProSim Project, no.: 559). We would like to thank Sonnhild Namingha from the Fraunhofer Institute Experimental Software Engineering for reviewing the first version of the article.